# An Effective Fusion Technique of Cloud Computing and Networking Series

K.Saravanan[1] and S.Akshaya[2], R.Pavithra[3] and K.Pushpavalli[4]

[1, 2] Asst. Professor/CSE, Erode Sengunthar Engineering College, Erode, India

[3, 4] B.E. C.S.E. Student, Erode Sengunthar Engineering College, Erode, India

Email:saravanankumarasamy@gmail.com,eternal.aks@gmail.com

*Abstract*— Cloud computing is making it possible to separate the process of building an infrastructure for service provisioning from the business of providing end user services. Today, such infrastructures are normally provided in large data centres and the applications are executed remotely from the users. One reason for this is that cloud computing requires a reasonably stable infrastructure and networking environment, largely due to management reasons. Networking of Information (NetInf) is an information centric networking paradigm that can support cloud computing by providing new possibilities for network transport and storage. It offers direct access to information objects through a simple API, independent of their location in the network. This abstraction can hide much of the complexity of storage and network transport systems that cloud computing today has to deal with. In this paper we analyze how cloud computing and NetInf can be combined to make cloud computing infrastructures easier to manage, and potentially enable deployment in smaller and more dynamic networking environments. NetInf should thus be understood as an enhancement to the infrastructure for cloud computing rather than a change to cloud computing technology as such. To illustrate the approach taken by NetInf, we also describe how it can be implemented by introducing a specific name resolution and routing mechanism.

## I. INTRODUCTION

The topic of Cloud Computing is gaining more and more attention in the service research community. The main idea is to make applications available on flexible execution environments primarily located in the Internet. Several flavours are known, and three important ones are depicted in the figure below.

Infrastructure as a service refers to the sharing of hardware resources for executing services, typically using virtualization technology. With this so-called Infrastructure as a Service (IaaS) approach, potentially multiple users use existing resources. The resources can easily be scaled up when demand increases, and are typically charged for on a per-pay-use basis. In the Platform as a Service (PaaS) approach, the offering also includes a software execution environment, such as an application server. In the Software as a Service approach (SaaS), complete applications are hosted on the Internet so that e.g. your word processing software isn't installed locally on your PC anymore but runs on a server in the network and is accessed through a web browser.

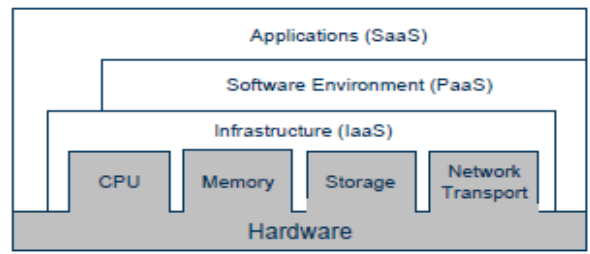

Figure 1. Cloud Computing Overview

At the same time, the networking research community is working on exploring the benefits arising from the new paradigm of content-centric or information-centric networking. Traditional networking architectures for the PSTN and the Internet solve the problem of connecting terminals or hosts in order to support a specific application such as telephony or WWW. To this end, traditional naming and addressing schemes employ box- or domain-oriented identities such as E.164 numbers for telephony, or IP addresses and URLs for the Internet. However, the end user is typically interested in reaching a destination object that sits behind or in the host, such as a human being or a file, rather than communicating with the host itself. As the destination objects move to new hosts, the host- or networkdependent identities of these objects must be updated.

Information-centric networking provides a solution to these issues by directly addressing the information objects instead of using the host-dependent or domain-dependent addressing schemes.

While URLs are also used to identify information objects, there is an important difference to how NetInf names information objects. URLs include the domain name or locator of the host at which the target object is stored and are therefore based on the traditional location-oriented communication paradigm. Consequently, links based on such URLs break when the host of a target object moves to a new location, or when the address of the host of the object changes. In the current Internet, there are several fixes to circumvent this problem, such as HTTP redirects and dynamic DNS. The location-independent object naming scheme of NetInf avoids this problem altogether, as the NetInf object names remain the same independent of any topology events, including location updates.

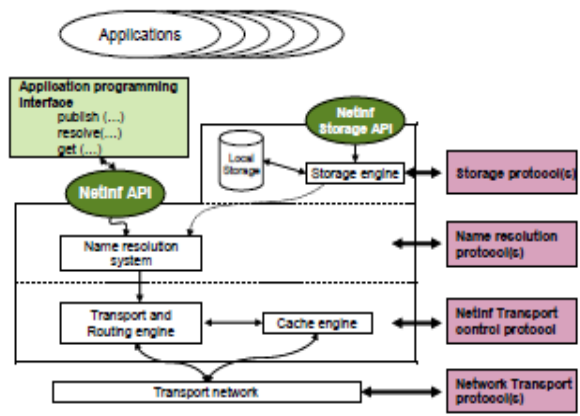

Figure 2. Networking of Information Overview

The basic idea of Network of Information (NetInf) is to move from today's host centric networking paradigm to an information centric networking paradigm where the information objects are the primary components of networking. One key aspect is that the information objects are named independent of the hosts they are stored on. The basic design ideas of NetInf are described in [1]. In NetInf, users request the information through a well-defined API by specifying the name of an information object using the get() primitive at the NetInf API in Figure 2. The name can have cryptographic properties, e.g. a hash of the content of a file can be part of the name. The name of the information object can then be used to verify the authenticity of the file. How and from where the object is retrieved is decided by the NetInf Transport and Routing engine. This makes it possible to react to changes in the network, both in terms of topology and load situation, in a flexible way. Information can be stored at arbitrary caches in the network (for short-term optimization, cf. Caching engine in Figure 2) and storage units (for long-term persistent storage, cf. Storage engine in Figure 2). It can also be retrieved from NetInf hosts that have already received the information and have stored it in their local cache. The NetInf Name Resolution System constitutes a flexible and scalable mechanism for handling the bindings between the object names and location, e.g. to support host, user and object mobility. From a cloud computing point of view, the network of information therefore offers new technology for dealing with the "networking resources" and "storage" boxes in Figure 1.

In this paper, we consequently discuss the cross-section between the cloud computing and the information centric paradigm. The remainder of the paper is structured as follows. In section II we discuss how cloud computing can benefit from Networking of Information (NetInf). Then in section III we make a first proposal for an architecture for putting cloud computing and NetInf together. We then go on to describe, in some more detail, the techniques that make NetInf able to provide the needed functionality. Finally we provide a conclusion and discuss further work.

## II. HOW CLOUD COMPUTING CAN BENEFIT FROM NETINF

Cloud computing is today offering an efficient environment for quick deployment of new services. In particular, it offers unrivalled opportunities for quickly scaling up the capacity for services that suddenly become popular. Infrastructure as a Service (IaaS) and Platform as a Service (PaaS) approaches have proven to be effective means for separating the service deployment from the provisioning of the required service infrastructure. The latter obviously is a much slower process as it includes hardware installation and possibly extended physical facilities, as well as establishment of new communication infrastructure. One of the major hurdles with current cloud computing is the management costs of the computing, storage and networking facilities [3]. To make the management tractable, today's approach to cloud computing depends on the site infrastructure and the networking topology being relatively stable.

The main benefit of what is presented in this paper is not a way to improve the cloud computing functionality as such by use of NetInf. We rather argue that by basing the cloud computing infrastructure on NetInf, many of today's issues with management of cloud computing infrastructures can be eased.

In this light, distributing cloud computing service objects across a more dynamic networking environment like moving networks, or a home networking environment, is a very difficult task with today's cloud computing technology. In addition, cloud computing service objects need to be copied between servers, e.g. to realize load sharing. When using existing technologies, such issues would need to be handled by a combination of more or less automated network management tools, routing protocols, and add-on protocols for host and site multihoming, as well as host and network mobility. This adds substantial complexity to the system, and the mechanisms may interact with each other as well as with NATs and firewalls in an unpredictable fashion.

The following paragraphs illustrate how introducing NetInf technology can help addressing the above cloud computing challenges. We also propose other possible benefits that NetInf can bring to cloud computing.

The Network of Information paradigm allows for a new approach to the problem of dynamic network topologies.While cloud computing brings about revolutionary technology to resource sharing, the Network of Information is a new approach to accessing information (in its widest sense) on a network. It brings a new abstraction to the networking layer: the notion of information objects.Compared to existing networking paradigms, this new approach is designed for directly accessing the information, rather than addressing it indirectly via the host or network domain containing the information. The location of the information thereby becomes secondary, which makes it much easier to deal with configuration changes and mobility. The addressing used in NetInf fully relies on naming the information itself, and not the location or network domain it is retrieved from. For a cloud computing platform, this significantly simplifies the way data is handled. This property also allows for hiding reconfigurations of the supporting nodes and networks from the cloud computing platform. It could potentially also make cloud computing deployable on a smaller scale, like home networking, and even in environments with mobile nodes and networks.

When the access of information objects is no longer done in an application-specific fashion, integration and composition of applications become easier, since they all access information using the same naming scheme. Access control is a topic of its own right in the NetInf work, but outside the scope of this paper.

The media distribution capabilities of NetInf provide a part of implementing IaaS. A part of IaaS or PaaS offerings is often the capability to efficiently distribute large amounts of content, e.g. video files. This can be realized via content distribution overlays, such as the well-known Akamai CDN [5]. Content distribution functionality is provided natively in a NetInf-enabled network, realized by the functions introduced in the previous section. Information is simply requested by specifying its name, while a NetInfenhanced networking layer takes care of choosing the optimal distribution mechanism, including caching and source selection. Similar to the popular peer-to-peer systems, clients can also serve as a new source for already downloaded information.

NetInf functionality will also support services and service composition. Without aiming to become a service platform in itself, the idea is that services can be treated as a type of information objects. Also a service needs to be unambiguously identified, and can have multiple instances in the network. Meta-data can be used to describe the characteristics of the service, and to support the proper selection of a service. This applies both to cases where the selection takes place manually, e.g. by a user browsing a service registry, and to cases where this happens in an automated fashion, e.g. in a service composition engine. In this way many mechanisms such as load balancing and mobility that are available through NetInf can be reused also for services Through a close cooperation of cloud computing and networking of information, network nodes and network resources, including various storage systems, have the potential of becoming truly transparent to the applications and the users. The natural evolution of networking is thus to move from networking of nodes to networking of information objects. But for users to feel comfortable with the, initially appealing, idea of just dropping their information objects into the network, for storage, processing and distribution, there are a number of issues that need to be addressed and are currently being worked on in the 4WARD project. These include security issues like confidentiality, integrity, privacy and access rights. Also requirements on reliability and availability will pose challenges.In section IV we propose a mechanism designed to meet this challenge.

## III. ARCHITECTURE OVERVIEW FOR CLOUD COMPUTING WITH NETINF

In Figure 1 it is shown how the basic cloud computing service IaaS is composed of a set of virtualized resources, namely CPU, memory, storage and network transport. In the previous section, we have already explained how NetInf can offer an enhanced solution for network transport and storage.

But let's take one step back and look at the two main reasons to virtualize resources: one is security - to restrict access to resources in order to offer a 'virtual private' environment, the other is resource separation in order to avoid resource conflicts.

NetInf can offer an alternative to the security aspect of virtualization for two of the cloud computing resources, storage and network transport. The NetInf networking architecture inherently secures the access to information objects and network resources by cryptographic means. There is therefore no need to virtualize storage and network transport resources to ensure the access rights. This stems from the fact that NetInf abstracts away the boxes and links that are implementing these resources and secures the resources themselves. Thus, there is no need to virtualize these boxes and links from a security perspective.

The need to virtualize for resource separation may or may not be a problem depending on the network configuration. With the current trend of reduced prices, especially on storage but also on transmission links, this might be dealt with by traditional overprovisioning, which can be combined with active network management in order to ensure that new (real) resources are added before resource conflicts appear. NetInf in combination with advanced virtualization techniques like Vnet proposed in 4WARD can provide a simpler solution to this problem by virtualizing both storage and network transport resources [6]. One key advantage with Vnet compared to other virtualization techniques is that it also addresses the problem of virtualizing wireless network resources.

The NetInf architecture provides an API for communication between arbitrary types of information objects, independently of which hosts they are attached to, and of how they move between hosts. By arbitrary information object we mean any information object which adheres to the NetInf naming scheme, and which is registered with the NetInf name resolution system. Examples of such objects are data files, service objects, or digital representations of physical objects, such as RFID tags. The NetInf API supports a mode of communication which is object-centric in the sense that only the object identity or a set of attributes are needed to access an object. The object naming scheme allows users and applications to create object names based on cryptographic hashes of the owner's public key This avoids the need for introducing a new naming authority.

Such object names are entirely independent of the location of the object. Using this naming scheme, the API hides the location of an object, as well as the dynamics of the underlying transport network. In addition, if several identical copies of an object exist in the network, the NetInf name resolution and routing system finds the "best"1 copy in an anycast fashion.

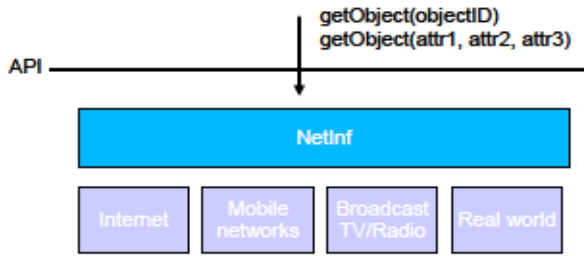

Figure 3. NetInf API

The API includes methods such as publish(objectName), resolve(objectName), and get(ObjectName). These methods allow for registrations of objects in the name resolutions system, resolving the location of an object, and establishing connectivity with the object. The API also includes methods for object storage and retrieval.

Figure 4 shows how NetInf can support a set of different cloud computing services. The cloud computing service uses NetInf objects, which communicate over a dynamic infrastructure network, such as a global network, using the NetInf API. The service uses object names to retrieve and interact with other objects over the API, and has no notion of object locators. The dynamic infrastructure network on the other hand uses traditional addresses (locators), and has no notion of service layer objects.

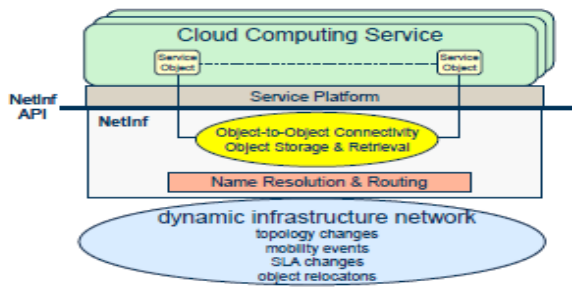

Figure 4. Architecture overview for the support of cloud computing services over NetInf.

A cornerstone of the NetInf architecture is the name resolution and routing system, which resolves the name of an object to a set of current network locators. The resolution mechanism is designed to handle highly dynamic network topologies in a scalable fashion, and provides an updated locator for a digital object that is moved between hosts, or for digital objects that are stored on mobile hosts, which in turn may be attached to moving networks. Also, the resolution mechanism is capable of handling multihoming of objects, hosts, and networks. Note that the capability of handling these rather general mobility and multihoming scenarios also provides a basis for the handling of dynamic events in fixed networks as described in section 2.1.

The name resolution and routing system is designed to scale to large networks and to a large number of objects (~$10^{15}$). Likewise, the routing system must allow for short convergence times also in a dynamic network topology. The focus of the next section is on the name resolution and routing system and its interoperation with the dynamic network infrastructure. A novel mechanism is described that allows for a strict separation between the object-centric view of the API on the one hand, and a highly dynamic network topology on the other hand.

### IV. NAME RESOLUTION AND OBJECT-TO-OBJECT ROUTING BASED ON LATE LOCATOR CONSTRUCTION

Traditional routing provides a network path between source and destination boxes that have locators such as IPv4 or E.164 addresses. The NetInf name resolution and routing system uses an object-to-object routing mechanism that provides a network path between any set of objects that can register with a network node. A key feature of the mechanism is that it can handle routing for a very large number of objects across a network that has a dynamic topology.

Below we describe an implementation of the NetInf name resolution and routing system based on the Late Locator Construction (LLC) mechanism. An overview of the locator construction procedure is shown in Figure 5.

The end-to-end routing problem is divided into three parts: (i) finding a path from the source object to the source core edge router at one edge of the core, (ii) Finding a path form the source core edge router through the core to the destination core edge router at another edge of the core, and (iii) finding a path from the destination edge router to the destination object.

LLC addresses the first and the third of these routing problems. To route through the core network, traditional routing can be used, since one of the requirements on the core network is that it is reasonably static. As the majority of the nodes will reside in the edge parts of the network, also a future core network should consist of much fewer nodes than today's internet and there should not be a problem to reuse today's routing for the foreseeable future.

With the LLC mechanism, the locator of an object is constructed (encoded) when packets are to be sent to or from the object in order to take fresh topology information into account. The object locator is thus constructed at the last moment before sending packets, which motivates the term late locator construction. By using fresh topology information when encoding a forwarding path in the locator, topology changes due to mobility or re-homing events can be taken into account.

To quickly construct locators that represent the current topology, LLC uses Attachment Registers (AR). Each node in the network is represented by an AR in the routing system. When the resolve() method of the NetInf

API is invoked, a global name resolution system based on a distributed hash table resolves the cryptographic ID of the target object into a locator of the AR that is associated with the object. Starting from this AR, the current locator of the object is constructed and returned.

The locator construction procedure described above is represented by steps 1-7 in Figure 5. The figure shows an object locator that describes a path from the core network to the object using a syntax with a sequence of attachments between adjacent network entities. The attachments are represented by the @ symbol. In a real implementation, the locator syntax would be more compact to reduce the overhead.

It should be noted that there exist various aggregation schemes that reduce the amount of mobility update signalling needed for mobile hosts and networks, for example the IETF Nemo protocol. However, such schemes are not readily applicable to the problem of arbitrary types of moving objects which is addressed by the NetInf architecture. Moreover, Nemo suffers from the wellknown pin-ball routing problem, i.e. the user data path traverses all the mobility agents of the moving networks involved in an end-to-end flow.

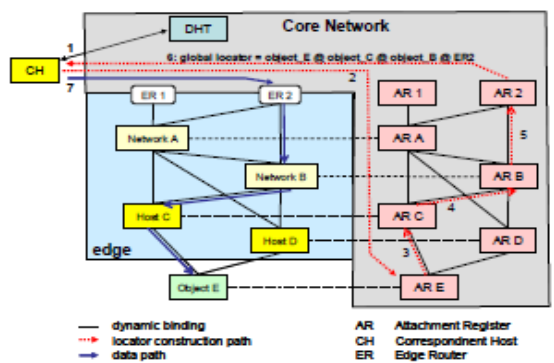

Figure 5. Overview of the Late Locator Construction procedure.

## V. CONCLUSION AND FUTURE WORK

The main purpose of this paper is to investigate what benefits the Networks of Information (NetInf) technology can bring to the infrastructure that is the foundation for cloud computing. In particular, the management of a cloud computing infrastructure can be simplified, as it does not have to deal with the details of storing and transporting information objects.

In this paper we have presented the NetInf architecture, and described how it can support cloud computing services by offering an API that hides the dynamics of object locations and network topologies. One single name resolution and routing mechanism is used, regardless of whether the dynamics depend on network reconfigurations, change of service level agreements, mobility events, rehoming events, or any other type of network event. The task of designing cloud computing services that are robust against object re-locations or changes in the topology of the underlying infrastructure network can thereby be significantly simplified.

To illustrate the NetInf approach, a novel routing mechanism based on late locator construction has been described that performs object-to-object routing rather than traditional host-to-host routing. This mechanism can operate over a highly dynamic network topology and allows for scalable handling of a very large number of objects.

Future work includes more detailed investigations on how NetInf can handle services, including the use of NetInf as a service directory for Web Services. Apart from the features described in this paper, also automated and distributed processing of information objects shall be investigated, e.g. to offer a delay-sensitive service as close as possible to the end-user. As both NetInf and virtualization (Vnet) of network resources are part of a common architecture being developed in the 4WARD project, we are also investigating which additional benefits their combination can bring to cloud computing.

REFERENCES

[1] Bengt Ahlgren, Matteo D'Ambrosio, Christian Dannewitz, Marco Marchisio, Ian Marsh, Börje Ohlman, Kostas Pentikousis, René Rembarz, Ove Strandberg, and Vinicio Vercellone. Design Considerations for a Network of Information (position paper). InReArch'08 - Re-Architecting the Internet, Madrid, Spain, December 9, In conjunction with ACM CoNEXT 2008.

[2] V. Devarapalli, R. Wakikawa, A. Petrescu, and P. Thubert. IETF RFC 3963, Network Mobility (nemo) Basic Support Protocol, January 2005.

[3] Rao Mikkilineni. Cloud Computing and the lLessons from the Past. 2009.

[4] Börje Ohlman, et al. First Netinf Architecture Description, fp7-ict- 2007-1-216041-4ward / D6.1. Technical report, January 2009.

[5] Kostas Pentikousis. Distributed Information Object Resolution. In Proc. Eighth International Conference on Networks (ICN), Gosier, Guadeloupe/France, March 2009. IEEE Computer Society Press.

[6] N. Egi, A. Greenhalgh, M. Handley, M. Hoerdt, F. Huici, and L. Mathy, "Towards High Performant Virtual Routers on Commodity Hardware", ACM CoNEXT, Madrid, Spain, December 2008.

[7] EU FP7 4WARD project, http://www.4ward-project.eu/

Authors

**Mr. Saravanan Kumarasamy** received the M.E degree 2008 in computer science and Engineering from Anna University, Chennai, India. He is currently working as an Assistant Professor at the Faculty of Engineering, Erode Sengunthar Engineering College, Erode, Tamilnadu. He has published 6 paper in International Journal, 07 papers in National Conference and 02 papers in International Conference. His current research interests are information security, Knowledge Mining and Computer Communications and DDoS Attacks.

He is currently pursuing Ph.D. under Anna University of Technology, Coimbatore.

**Ms. Akshaya Subramani** received the M.E degree 2012 in Computer science and Engineering from Karpagam University, Coimbatore, India. She is currently working as an Assistant Professor at the Faculty of Engineering, Erode Sengunthar Engineering College, Erode, Tamilnadu. He has published 02 papers in National Conference Her current research interests are information security, Cryptography.